\documentclass[preprint]{raa}            
\usepackage{graphicx,times}             
\usepackage{natbib}
\usepackage{adjustbox}
\usepackage[version=4]{mhchem}
\usepackage{amssymb,amsmath}
\bibpunct{(}{)}{;}{a}{}{,}
\footnotesize
\newdimen\digitwidth
\setbox0=\hbox{\rm0}
\digitwidth=\wd0
\catcode`!=\active
\def!{\kern\digitwidth}
\normalsize

\usepackage[colorlinks=true, allcolors=blue]{hyperref}

\begin{document}

\title{ALMA and GMRT Studies of Dust Continuum Emission and Spectral Lines Toward  Oort Cloud Comet C/2022 E3 (ZTF)}

\volnopage{Vol.0 (20xx) No.0, 000--000}      
\setcounter{page}{1}          

\author{
Arijit Manna\inst{1}
\and Sabyasachi Pal\inst{1}
\and Sekhar Sinha\inst{2}
\and Sushanta Kumar Mondal\inst{2}
}

\institute{
Department of Physics and Astronomy, Midnapore City College, Paschim Medinipur 721129, India\\
\and Department of Physics, Sidho-Kanho-Birsha University, Ranchi Road, Purulia 723104, India\\
\vs\no
{\small Received 20xx month day; accepted 20xx month day}}

\abstract{
The atomic and molecular compounds of cometary ices serve as valuable knowledge into the chemical and physical properties of the outer solar nebula, where comets are formed. From the cometary atmospheres, the atoms and gas-phase molecules arise mainly in three ways: (i) the outgassing from the nucleus, (ii) the photochemical process, and (iii) the sublimation of icy grains from the nucleus. In this paper, we present the radio and millimeter wavelength observation results of Oort cloud non-periodic comet C/2022 E3 (ZTF) using the Giant Metrewave Radio Telescope (GMRT) band L and the Atacama Large Millimeter/Submillimeter Array (ALMA) band 6. We do not detect continuum emissions and an emission line of atomic hydrogen (H{\sc I}) at rest frequency 1420 MHz from this comet using the GMRT. Based on ALMA observations, we detect the dust continuum emission and rotational emission lines of methanol (\ce{CH3OH}) from comet C/2022 E3 (ZTF). From the dust continuum emission, the activity of dust production (Af$\rho$) of comet ZTF is 2280$\pm$50 cm. Based on LTE spectral modelling, the column density and excitation temperature of \ce{CH3OH} towards C/2022 E3 (ZTF) are (4.50$\pm$0.25)$\times$10$^{14}$ cm$^{-2}$ and 70$\pm$3 K. The integrated emission maps show that \ce{CH3OH} was emitted from the coma region of the comet. The production rate of \ce{CH3OH} towards C/2022 E3 (ZTF) is (7.32$\pm$0.64)$\times$10$^{26}$ molecules s$^{-1}$. The fractional abundance of \ce{CH3OH} with respect to \ce{H2O} in the coma of the comet is 1.52\%. We also compare our derived abundance of \ce{CH3OH} with the existence modelled value, and we see the observed and modelled values are nearly similar. We claim that \ce{CH3OH} is formed via the subsequential hydrogenation of formaldehyde (\ce{H2CO}) on the grain surface of comet C/2022 E3 (ZTF).
}

\authorrunning{Manna et al.}            
\titlerunning{Continuum and spectral lines in C/2020 E3 (ZTF)}  

\maketitle

\section{Introduction}
\label{sec:intro} 
Comets are kilometre-sized, ice-rich objects made of refractory, and volatile chemical compounds \citep{eh00}. This is the most pure remnant of the formation of the solar system, just 4.6 billion years ago. Comets are the most excellent insight to study the volatile compositions of the solar protoplanetary disks since they gather some of the solar system's most ancient and primordial substances, including ices. Previous studies show that water (\ce{H2O}) and different organic matters may have reached the early Earth through comets as well \citep{har11}. Since comets travel in the interstellar medium (ISM), they collect complex compounds, including amino acids (R--CH(NH$_{2}$)--COOH), from molecular clouds, star-formation regions, and protoplanetary disks and transport them into the planetary atmosphere via collision with the planet \citep{eh00}. Thus, it is crucial to understand the variability in composition and isotopic ratios of cometary material to evaluate such scenarios \citep{at03, bo15}. Previously, more than 60 molecular compounds were identified towards comet 67P/Churyumov-Gerasimenko between 2014 and 2016 using the Rosetta Orbiter Spectrometer for Ion and Neutral Analysis (ROSINA), and their chemical compositions have been investigated using both in situ and remote ways (see \cite{biv19} and references therein). In recent years, high-sensitivity ground-based millimeter wavelength telescopes have identified over 25 complex molecular compounds from the cometary atmospheres (see Figure~1 of \cite{biv19} and references therein). Earlier, \cite{al16} detected the simplest amino acid glycine (\ce{NH2CH2COOH}), phosphorus (P), and different complex compounds from the 67P/Churyumov-Gerasimenko using the ROSINA. After that, \citet{had19} made a chemical model and showed that \ce{NH2CH2COOH} ejected from the nucleus due to the sublimation of ice from the dust particles. Recently, centimeter and millimeter wavelength observations found evidence of atomic hydrogen (HI) and molecular emission lines of HCN, HNC, \ce{CH3OH}, CS, \ce{CH3CN}, \ce{H2CO}, SO, \ce{HC3N}, and \ce{H2S} from comet C/2020 F3 (NEOWISE) \citep{biv22, pal24}.

\begin{figure*}
\centering
\includegraphics[width=1.0\textwidth]{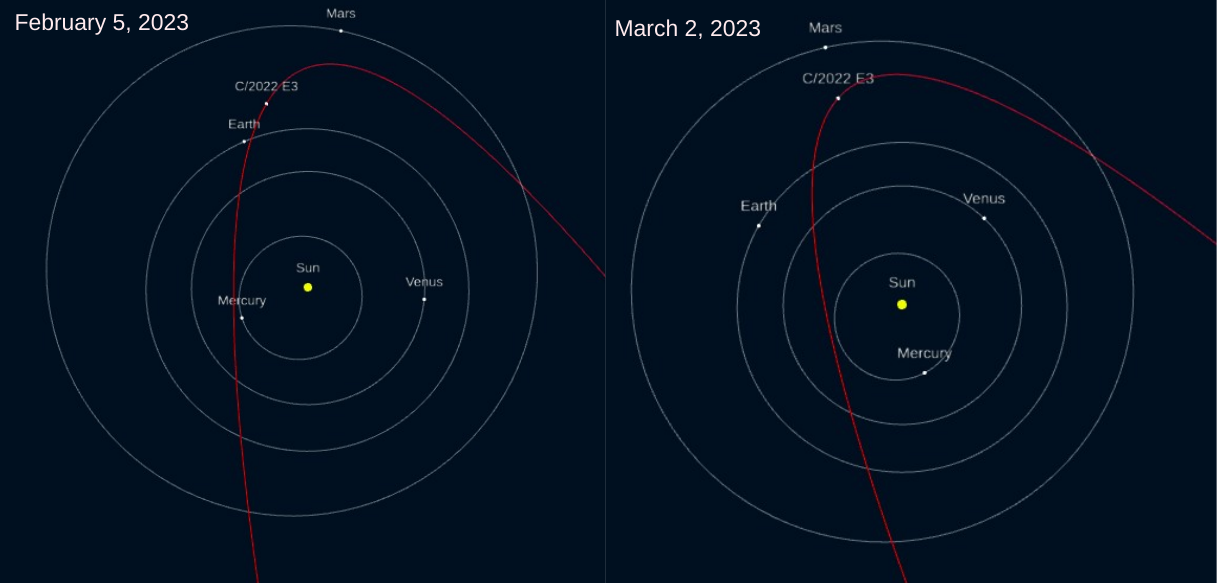}
\caption{Orbital configuration of comet C/2022 E3 (ZTF) and inner solar system on February 5, 2023 (date of GMRT observation) and March 2, 2023 (date of ALMA observation).} 
\label{fig:track}
\end{figure*}

One of the most common interstellar molecules, methanol (\ce{CH3OH}), was discovered in space for the first time over 54 years ago \citep{bal70}. In the ISM, \ce{CH3OH} has been found in gaseous and solid forms \citep{eh00}. The evidence of \ce{CH3OH} is found in hot molecular cores, hot corinos, dark clouds, the Galactic Centre, and protoplanetary disks \citep{her09, th17, man20, boo21, chan22}. Earlier, \cite{dra19} claimed the detection of \ce{CH3OH} with transition J = 5(1,1)--4(1,1) towards Saturn's moon Enceladus using IRAM 30-m telescope, but a single transition line of complex molecules does not provide a secure detection. So it is challenging to believe the detection of \ce{CH3OH} towards Enceladus by \cite{dra19}. The emission lines of \ce{CH3OH} are also found in the cometary atmosphere. In the solar system, the \ce{CH3OH} lines were first found from comets Austin (1990 V) and Levy (1990 XX) with the IRAM 30-m telescope \citep{boc94a}. The emission lines of \ce{CH3OH} are also found from comets P/Swift-Tuttle (1992) \citep{boc94b}, Hyakutake (C/1996 B2) \citep{wo97}, Lee (C/1999 H1) \citep{biv00}, HaleBopp (C/1995 O1) \citep{biv99}, P/Halley \citep{eb94}, C/2012 S1 (ISON) \citep{ag14}, C/2013 R1 (Lovejoy) \citep{ag14}, C/2012 F6 (Lemmon) \citep{bo17}, C/2012 K1 (PanSTARRS) \citep{co17}, 46P/Wirtanen \citep{co23}, and C/2020 F3 (NEOWISE) \citep{biv22}. Previously, \citet{gr82} demonstrated that the chemical composition of cometary grains exhibits a strong similarity to those found in the ISM and circumstellar disks, which indicates the formation pathways of \ce{CH3OH} towards the cometary atmosphere are nearly similar to interstellar grains. Earlier, \cite{gar19} showed that the formation pathways of \ce{CH3OH} towards the cometary atmosphere are similar to the star-formation regions.

\begin{table*}
\caption{Orbital properties of comet ZTF.}	
\centering
\begin{adjustbox}{width=0.9\textwidth}
\begin{tabular}{cccccccccc}
\hline 
Parameters &Symbol&Value\\
\hline
Orbit eccentricity& e&1.0000290\\	
Orbit inclination& i&109.16920$^{\circ}$\\
Perihelion distance&q&1.1126880 AU\\
Aphelion distance&Q&2800 AU\\
Semi-major axis&a&$-$4087 AU\\
Orbital period&P&50,000 years\\
Time of perihelion passage&T&2459957.18000 JD\\
Longitude of ascending node&$\Omega$&302.53990$^{\circ}$\\
Argument of perihelion&$\omega$&145.80970$^{\circ}$\\
Longitude of perihelion&L&315.11506$^{\circ}$\\
Latitude of perihelion&B&32.05853$^{\circ}$\\
Classification& &Nearly isotropic (a $>$ 10000 AU)\\ 
\hline
\end{tabular}
\end{adjustbox}\\
\label{tab:orbitpar}
Note: The orbital parameters of comet ZTF are taken from \citet{bo22} and NASA JPL Horizons System.
\end{table*}
	
C/2022 E3 (ZTF) (hereafter ZTF) is a long-period Oort-cloud comet (initial semi-major axis of 2000 AU) that was discovered on March 2, 2022, using the Zwicky Transient Facility at a heliocentric distance of 4.3 AU \citep{bo22}. The orbital properties of comet ZTF are shown in Table~\ref{tab:orbitpar}. This comet came closest to Earth on February 1, 2023, at a distance of 0.28 AU. The close distance indicates that comet ZTF is a near-Earth object. In March 2022, comet ZTF showed interesting cometary features, including a green coma of diatomic carbon (\ce{C2}), a yellow dust tail, and a thin ion tail. After the identification of this comet, \citet{liu24} observed that the heliocentric distance of ZTF decreased with time, which indicates the sublimation of the volatile species from the grain surface of the nucleus due to the acceleration of solar heat, which enhances the luminosity of green coma. The production rates of OH and \ce{H2O} towards comet ZTF were $Q_{OH}$ = 3.51$\times$10$^{28}$ s$^{-1}$ and $Q_{H_{2}O}$ = 4.8$\times$10$^{28}$ s$^{-1}$, respectively \citep{je22, sc23}. The production rates of CN, \ce{C3}, and \ce{C2} on 10 March 2023 were (5.43$\pm$0.11)$\times$10$^{25}$ s$^{-1}$, (2.01$\pm$0.04)$\times$10$^{24}$ s$^{-1}$, and (3.08$\pm$0.50)$\times$10$^{25}$ s$^{-1}$, respectively \citep{bol24}. \citet{liu24} shows that the size of the nucleus of the comet ZTF is 2.79 km. In February 2023, the Trivandrum Observatory post-perihelion observation data showed that the apparent sizes of the tail and head of the comet ZTF were noticeably decreased \citep{joy23}. Recently, \cite{biv24} detected several rotational emission lines of complex organic molecules, including \ce{CH3OH}, towards comets ZTF and C/2021 A1 (Leonard) using the IRAM 30-m telescope. \cite{biv24} also claimed the detection of ethylene glycol ((\ce{CH2OH})$_{2}$) and methyl cyanide (\ce{CH3CN}) from comets ZTF and C/2021 A1 (Leonard). \cite{biv24} also does not discuss the formation pathways of \ce{CH3OH} towards comet ZTF.
	
This paper presents the centimeter and millimeter wavelength observations of comet ZTF using the high-resolution GMRT and ALMA. We also detected the rotational emission lines of \ce{CH3OH} from the coma region of this comet. The observational methods and data analysis techniques used in this study are described in Section~\ref{obs}. The results and discussions of the detection of dust continuum and line emissions are shown in Section~\ref{res}. The conclusions are presented in Section~\ref{conclu}.

\begin{figure*}
\centering
\includegraphics[width=1.0\textwidth]{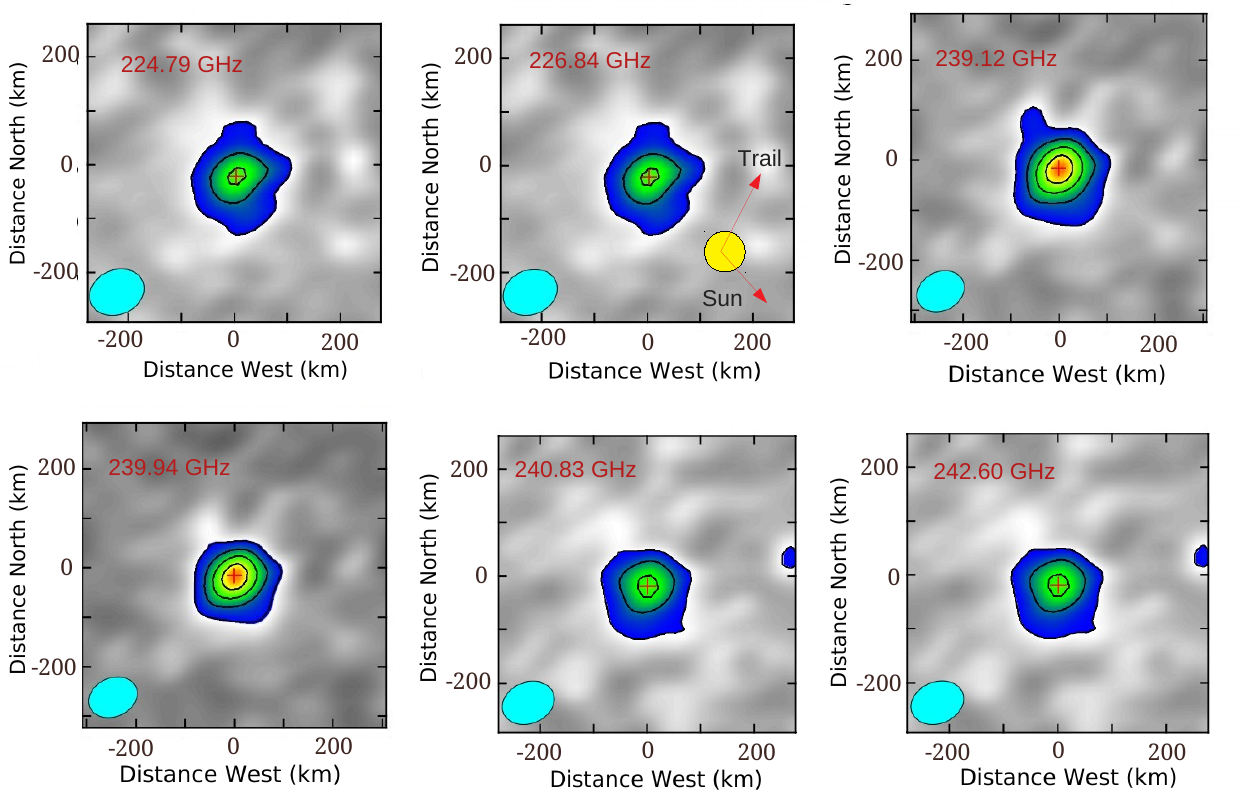}
\caption{Millimeter-wavelength dust continuum images of the comet ZTF. The direction of the Sun and the motion of the comet are shown in the first-panel image. The cyan-coloured circles are the synthesized beam of the dust continuum images. The contour levels start at 3$\sigma$ and increase by a factor of $\surd$2. The cross sign is the comet position, which is provided by the NASA JPL Horizon.}
\label{fig:continuum}
\end{figure*}

\section{Observations and data reductions}
\label{obs}
\subsection{Atacama Large Millimeter/Submillimeter Array (ALMA)}
We used the publicly available archival data of comet ZTF, which was observed on March 2, 2023, using the high-resolution Atacama Large Millimeter/Submillimeter Array (ALMA) 12-m arrays with band 6 (PI: Martin Cordiner, ID: 2022.1.00997.T). This observation aimed to study the millimeter-wavelength dust continuum emission and detection of volatile molecules. On March 2, 2023, the distance between the comet ZTF and Earth was 0.96 AU. Similarly, the distance between ZTF and the Sun was 1.35 AU. The orbital configuration of comet ZTF on March 2, 2023, is shown in Figure~\ref{fig:track}. This observation was carried out in the frequency ranges of 223.85--225.73 GHz, 225.91--227.79 GHz, 239.01--239.25 GHz, 239.89--240.01 GHz, 240.78--240.90 GHz, and 241.67--243.55 GHz with a spectral resolution of 976.56 kHz. A total of 39 antennas were used during the observation. The minimum and maximum baselines were 15.3 m and 783.1 m. The on-source integration time was 44.352 m. During the observation, the comet was continuously tracked, and the phase centre position in the sky was updated in real-time using the JPL Horizons orbital solution \#JPL 81. During the observation of comet ZTF, the flux and bandpass calibrator were taken as J0423--0120. Similarly, J0442--0017 was used as a phase calibrator.
	
For the analysis of the raw data of ZTF, we used the Common Astronomy Software Application (CASA 5.4.1) with standard pipeline scripts delivered by the Joint ALMA Observatory (JAO) \citep{te22}. For flux calibration, we used the task SETJY with the Perley-Butler 2017 flux calibrator model \citep{per17}. We also used the pipeline tasks {HIFA\_BANDPASSFLAG} and {HIFA\_FLAGDATA} to remove the bad antenna data. After analysing the raw data, we split the target source (ZTF\_C2022\_E3) with the help of the CASA task MSTRANSFORM. Before imaging, we subtract the continuum from the visibilities by fitting the second-order polynomial to the line-free channels in all spectral windows. We also used the CASA task CVEL for Doppler correction to the rest frame of the comet. We make the dust continuum and spectral images of ZTF using the task TCLEAN with hogbom deconvolution and the SPEC = CUBE parameter. To make the dust continuum emission images, we used line-free channels. At last, we corrected the primary beam pattern in both continuum and spectral images using the task IMPBCOR.
	
\begin{table}{}
\centering
\caption{Dust continuum emission properties of comet ZTF based on ALMA data.}
\begin{tabular}{cccccccc}
\hline
Frequency&Integrated flux& Peak flux&RMS &Synthesized beam &Deconvolved source size\\
(GHz)		&   ($\mu$Jy) &  ($\mu$Jy beam$^{-1}$) &($\mu$Jy) &($^{\prime\prime}$) &($^{\prime\prime}$)\\                   
\hline
224.79&124$\pm$17&120$\pm$10&9.45&0.53$^{\prime\prime}$$\times$0.42$^{\prime\prime}$ &0.51$^{\prime\prime}$$\times$0.44$^{\prime\prime}$ \\
			
226.84&132$\pm$49&106$\pm$12&5.87& 0.53$^{\prime\prime}$$\times$0.41$^{\prime\prime}$   &0.49$^{\prime\prime}$$\times$0.42$^{\prime\prime}$\\
			
239.12&138$\pm$17&109$\pm$13&7.52&0.50$^{\prime\prime}$$\times$0.39$^{\prime\prime}$    &0.47$^{\prime\prime}$$\times$0.38$^{\prime\prime}$\\
			
239.94&140$\pm$25&112$\pm$9&5.20&0.50$^{\prime\prime}$$\times$0.39$^{\prime\prime}$    &0.48$^{\prime\prime}$$\times$0.38$^{\prime\prime}$\\
			
240.83&145$\pm$38&115$\pm$10&4.68&0.50$^{\prime\prime}$$\times$0.39$^{\prime\prime}$    &0.48$^{\prime\prime}$$\times$0.37$^{\prime\prime}$\\
			
242.60&155$\pm$32&120$\pm$15&6.68&0.50$^{\prime\prime}$$\times$0.39$^{\prime\prime}$    &0.49$^{\prime\prime}$$\times$0.38$^{\prime\prime}$\\
\hline
\end{tabular}
\label{tab:continuum}\\
\end{table}
	
\subsection{Giant Metrewave Radio Telescope (GMRT)}
Comet ZTF was also observed on February 5, 2023, using the Giant Meterwave Radio Telescope (GMRT) band L (1050--1450 MHz). GMRT is located in Khodad near Pune in India (PI: Arijit Manna, ID: ddtC267). This observation aimed to study the radio wavelength continuum emission in frequency ranges of 1050--1450 MHz and the emission/absorption line of atomic HI at 1420 MHz. On February 5, 2023, the distance between the ZTF and Earth was 0.30 AU. The distance between ZTF and the Sun was 1.16 AU on that day. The orbital configuration of comet ZTF on February 5, 2023, is shown in Figure~\ref{fig:track}. During the observation, a total of 30 fully steerable, 45-m-diameter antennas were used, with minimum and maximum baselines of $\sim$100 m and $\sim$25 km, respectively. The National Centre for Radio Astrophysics (NCRA), a division of the esteemed Tata Institute of Fundamental Research (TIFR), operates this radio telescope. A total of 8192 number of channels were used with a spectral resolution of 0.0488 MHz. The on-source integration time was 3 hours and 15 minutes. During the observation, quasar 3C 147 was used as a flux and phase calibrator.
	
We used the standard calibration method using CASA to reduce the data of comet ZTF. We import the raw GMRT data in CASA using the task IMPORTGMRT. During calibration, we remove the bad spectral channels and damage antenna data. To calibrate the flux, we used the SETJY task, utilizing the Perley-Butler 2017 flux calibrator model \citep{per17}. We also make the bandpass calibration using the 3C 147. After gain calibration and transferring the gain calibration to the target, we applied the task MSTRANSFORM to split the target data. We make the continuum emission images of ZTF with w-projection and the multi-frequency synthesis mode with 2nd-order expansion \citep{rau11}. We also used the GAINCAL and APPLYCAL tasks for better image sensitivity. After making the continuum emission images, we subtract the continuum emission from the visibility using the task UVCONTSUB. We make the spectral images using the TCLEAN task with the SPECMODE = CUBE parameter. We also applied the task WPBGMRT for primary beam correction of the continuum and spectral images.
	
\begin{figure*}
\centering
\includegraphics[width=1.0\textwidth]{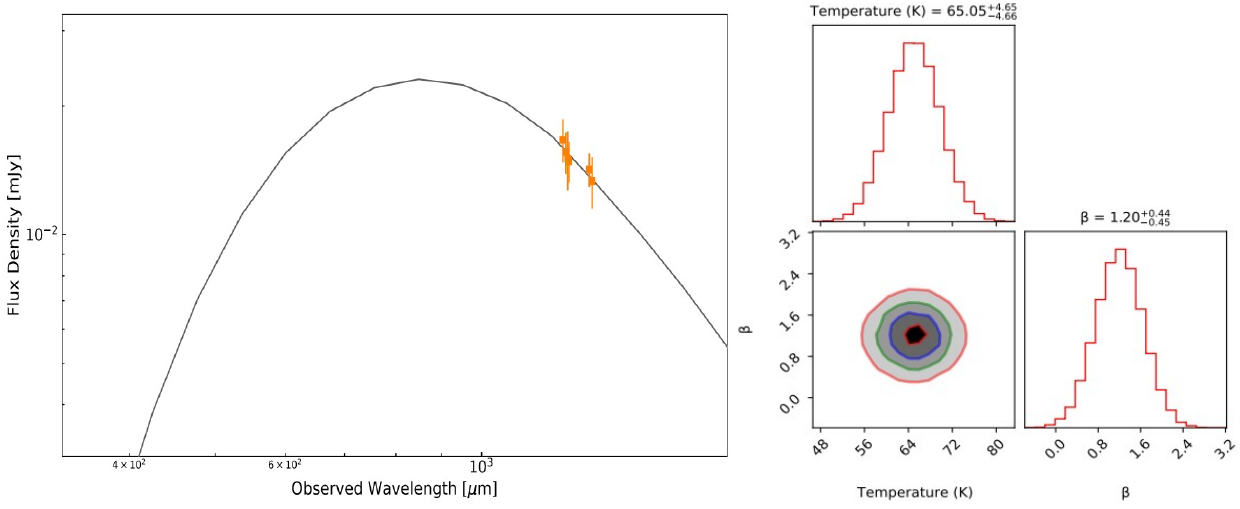}
\caption{SED plot of comet ZTF based on ALMA band 6 observations (left panel). In the SED plot, the solid orange lines are the fluxes of ZTF with error bars, and the black line indicates the best-fitted SED with the blackbody model. The right panel shows a corner diagram demonstrating the results of our MCMC parameter estimation for the SED model. The diagonal panels display 1-D histograms representing the marginalized posterior distributions for dust temperature and dust spectral index ($\beta$). Meanwhile, the off-diagonal panels illustrate 2-D projections of the posterior probability distributions, showcasing the correlations between each pair of parameters.}
\label{fig:sed}
\end{figure*}
		
\section{Result \& Discussion}
\label{res}
\subsection{Results based on ALMA data}
\subsubsection{Dust continuum emission towards comet ZTF}
We observed that the comet ZTF is detected in the frequency ranges of 224.79 GHz to 242.60 GHz, as shown in Figure~\ref{fig:continuum}. To estimate the dust continuum parameters of comet ZTF, we fitted the 2D Gaussian using the CASA task IMFIT over the cometary area with a circle of radius 1.7$^{\prime\prime}$. The physical parameters are shown in Table~\ref{tab:continuum}. The direction of motion of comet ZTF and the Sun is shown in Figure~\ref{fig:continuum}. Comet ZTF is not resolved in the frequency ranges of 224.79 GHz to 242.60 GHz as the deconvolved source sizes are smaller than the synthesized beam sizes. We plot the spectral energy distribution (SED) of the ZTF in the frequency ranges of 224.79 GHz to 242.60 GHz. After SED plotting, we fitted the comet flux values using the blackbody model with the help of the Python package ASTROPY (version 5.0). We used the Markov Chain Monte Carlo (MCMC) algorithm to fit the blackbody model over the observed flux values. The SED plot and corner diagram based on MCMC analysis are shown in Figure~\ref{fig:sed}. After SED analysis, we found the dust temperature and dust spectral index ($\beta$) of comet ZTF were 65.05$\pm$4.65 K and 1.20$\pm$0.44.
	
\begin{table*}
	\centering
	\caption{Comparision of the activity of dust production (Af$\rho$) with comet ZTF and other comets.}
	\begin{tabular}{lcccc}
		\hline
		Comet & Af$\rho$ & $R_h$ &Reference \\
		& (cm) & (AU)& \\
		\hline
		C/2023 E3 (ZTF)&2280$\pm$50&1.35&This study\\
		4P$/$Faye & 1146.0$\pm$1.2&1.62&\citet{gi24} \\
		6P$/$d$'$Arrest &$203.8\pm1.5$&1.54&\citet{gi24} \\
		11P$/$Tempel-Swift-LINEAR &   $17.8\pm0.3$&1.39&\citet{gi24} \\
		108P$/$Ciffreo & $146.2\pm1.5$&1.67&\citet{gi24} \\
		114P$/$Wiseman-Skiff &$76.7\pm0.3$ &1.59&\citet{gi24} \\
		132P$/$Helin-Roman-Alu &$128.6\pm0.5$ &1.70&\citet{gi24} \\
		156P$/$Russell-LINEAR &$681.6\pm0.4$ &1.35&\citet{gi24} \\
		254P$/$McNaught &  $360.0\pm6.5$ &3.68&\citet{gi24} \\
		398P$/$Boattini &$72.1\pm0.1$ &1.31&\citet{gi24} \\
		409P$/$LONEOS-Hill &$53.6\pm1.0$ &1.75&\citet{gi24} \\
		425P$/$Kowalski &$63.3\pm3.6$ &2.90&\citet{gi24} \\
		449P$/$Leonard (2020 S6) &$8.2\pm0.6$ &1.88&\citet{gi24} \\
		P$/$2019 LD2 (ATLAS) & $303.4\pm21.7$ &4.61&\citet{gi24} \\
		P$/$2020 T3 (PANSTARRS) &   $17.5\pm0.5$ &1.47&\citet{gi24} \\
		P$/$2020 U2 (PANSTARRS) &   $90.8\pm0.8$ &1.88&\citet{gi24} \\
		P$/$2020 WJ5 (Lemmon) & $193.4\pm16.1$ &5.07&\citet{gi24} \\
		P$/$2021 Q5 (ATLAS) &   $41.0\pm0.9$ &1.27&\citet{gi24} \\
		67P/Churyumov-Gerasimenko&233$\pm12$&1.36&\citet{sc06}\\
		P/Halley&20400$\pm$110&1.53&\citet{sch98}\\
		C/2014 UN$_{271}$ (Bernardinelli-Bernstein)&15000$\pm$250&20&\citet{le22} \\
		\hline
	\end{tabular}
	\label{tab:table1}
\end{table*} 
	
Now, we estimate the activity of dust production (Af$\rho$) of comet ZTF using the following equation \citep{ah84}:
\begin{equation} \label{equation1}
Af\rho = \frac{4 \, R_h^2 \, \Delta^2}{\rho}\frac{F_{com}}{F_\odot} 
\end{equation}
	
In equation~\ref{equation1}, $R_{h}$ is the distance between comet and Sun in AU, $\Delta$ is the distance between Earth and comet in cm, $F_{com}$ is the flux of comet in unit of erg cm$^{-2}$ s$^{-1}$, $\rho$ $\sim$ 6500 km is the nucleocentric distance of the comet ZTF \citep{bol24}, and $F_\odot$ is the solar flux in unit of erg cm$^{-2}$ s$^{-1}$ at 1 AU. The solar flux is taken from the high-resolution solar spectrum of \citet{ku84}. During the estimation of Af$\rho$, we used the flux value at frequencies of 242.60 GHz. Using equation 1, the calculated value of Af$\rho$ of comet ZTF, based on ALMA data, is 2280$\pm$50 cm, which is nearly similar to the estimated value of \cite{je23}. Recently, \citet{bol24} estimated the value of Af$\rho$ in optical wavelengths toward comet ZTF as 1483$\pm$40 cm, which is smaller than the estimated value of Af$\rho$ using ALMA data in millimeter wavelengths. Our estimated higher value of Af$\rho$ using the ALMA data indicates a high level of dust production due to the close distance between ZTF and the Sun during our observation.
	
Now, we compare the Af$\rho$ value of the comet ZTF with different dynamical types of other comets, as shown in Table~\ref{tab:table1}. After comparison, we observe that comet P/Halley exhibits a higher Af$\rho$ at $R_h$ of 1.53 AU. Similarly, we found the Af$\rho$ value of comet ZTF is 2280$\pm$50 cm at $R_h$ of 1.35 AU, which is higher than the rest of the comets but lower than the comet P/Halley and C/2014 UN$_{271}$ (Bernardinelli-Bernstein). We found different Af$\rho$ values for different comets owing to the different grain sizes, perihelion distances, and ice compositions, which produced more dust as ice vaporized and released dust particles.
	
\begin{figure*}
	\centering
	\includegraphics[width=1.0\textwidth]{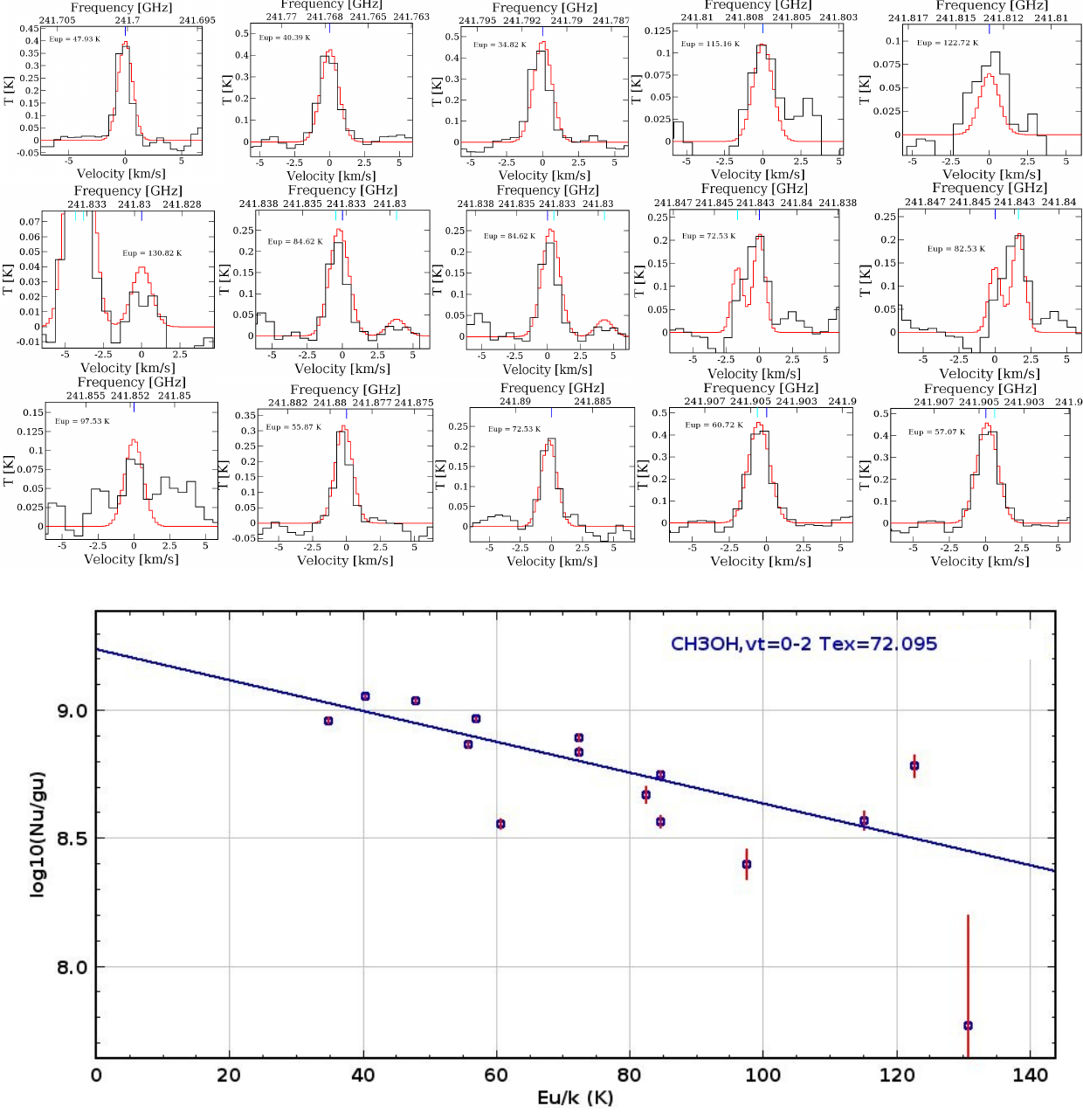}
	\caption{Rotational emission lines of \ce{CH3OH} towards comet ZTF (upper panel). In the spectra, the black lines are the observed spectra, and the red lines are the best-fit LTE model spectra of \ce{CH3OH}. The lower panel shows the rotational diagram of \ce{CH3OH}.}
	\label{fig:line}
\end{figure*}

\subsubsection{Methanol (\ce{CH3OH}) towards comet ZTF}
We extracted the molecular spectra from the spectral images by making a 0.8$^{\prime\prime}$ diameter circular region over the cometary area. The synthesized beam sizes of the spectral images in the frequency ranges of 223.85--225.73 GHz, 225.91--227.79 GHz, 239.01--239.25 GHz, 239.89--240.01 GHz, 240.78--240.90 GHz, and 241.67--243.55 GHz were 0.50$^{\prime\prime}\times$0.39$^{\prime\prime}$, 0.51$^{\prime\prime}\times$0.39$^{\prime\prime}$, 0.51$^{\prime\prime}\times$0.38$^{\prime\prime}$, 0.50$^{\prime\prime}\times$0.38$^{\prime\prime}$, 0.50$^{\prime\prime}\times$0.39$^{\prime\prime}$, and 0.51$^{\prime\prime}\times$0.39$^{\prime\prime}$. For spectral analysis, we used the CASSIS software \citep{vas15}. For detection of the rotational emission lines of \ce{CH3OH}, we used the Local Thermodynamic Equilibrium (LTE) model with the Cologne Database for Molecular Spectroscopy (CDMS) databases \citep{mu05}. To fit the local thermodynamic equilibrium (LTE) spectral model to the observed emission lines of \ce{CH3OH}, we utilized the MCMC algorithm within the CASSIS software package. After spectral analysis, we detected a total of 15 transition lines of \ce{CH3OH} from the comet ZTF. The detected emission lines of \ce{CH3OH} exhibited upper state energies ($E_{up}$) ranging from 34.82 K to 130.82 K. \ce{CH3OH} is a slightly asymmetric top molecule with an internal torsional motion of \ce{CH3} around the molecular symmetry axis relative to the OH radical. The transitions of \ce{CH3OH} are explained by $J$, $\pm$$K_{a}$, and $\Gamma$. In transitions of \ce{CH3OH}, we used $\Gamma$ = A, where $\Gamma$ is associated with `--' or `+’ sign indicates the parity. Similarly, we also used the $\Gamma$ = E in the transitions of \ce{CH3OH} because, in the absence of parity, the energy levels (E1 and E2) of \ce{CH3OH} are distinguished by $\pm$$K$. After detection of the rotational emission lines of \ce{CH3OH}, we obtained molecular transitions (${\rm J^{'}_{K_a^{'}\Gamma^{'}}}$--${\rm J^{''}_{K_a^{''}\Gamma^{''}}}$), upper state energy ($E_u$) in K, Einstein coefficients ($A_{ij}$) in s$^{-1}$, line intensity ($S\mu^{2}$) in Debye$^{2}$, full-width half maximum (FWHM) in km s$^{-1}$, optical depth ($\tau$), and integrated intensities ($\rm{\int T_{mb}dV}$) in K km s$^{-1}$. The detected emission lines and spectral line parameters of \ce{CH3OH} are shown in Figure~\ref{fig:line} and Table~\ref{tab:MOLECULAR DATA}. We also observed that the detected emission lines of \ce{CH3OH} are optically thin. After spectral line analysis using the LTE model, we found the column density and excitation temperature of \ce{CH3OH} are (4.50$\pm$0.25)$\times$10$^{14}$ cm$^{-2}$ and 70$\pm$3 K respectively. During LTE fitting, we used the source size $\sim$0.5$^{\prime\prime}$. To verify the temperature of \ce{CH3OH}, we also made the rotational diagram of \ce{CH3OH}, which is shown in Figure~\ref{fig:line}. The methods and theory of the rotational diagram are well discussed in \cite{man24New}. After the rotational diagram, we found the rotational temperature of \ce{CH3OH} is 72$\pm$12 K, which is very close to the excitation temperature of \ce{CH3OH}. Recently, \cite{biv24} also claimed the detection of \ce{CH3OH} towards comet ZTF, and they estimated the rotational temperature of \ce{CH3OH} is 58$\pm$4 K using the rotational diagram. The derived rotational temperature of \ce{CH3OH} by \cite{biv24} is lower than that of the present paper. The estimated rotational temperature of \ce{CH3OH} by \cite{biv24} is less because it is measured by the IRAM 30 m single dish, whose resolution is much less than the present observation. \cite{biv24} could not study the spatial distribution of \ce{CH3OH} towards ZTF. 

\begin{table*}
\centering
\scriptsize 
\caption{Spectral line properties of \ce{CH3OH} towards comet ZTF.}
\begin{adjustbox}{width=1.0\textwidth}
\begin{tabular}{ccccccccccccccccccc}
\hline 
Frequency &Quantum number&$E_{u}$ & $A_{ij}$ &g$_{up}$&$S\mu^{2}$&FWHM&$\rm{\int T_{mb}dV}$&Optical depth\\
(GHz)&(${\rm J^{'}_{K_a^{'}\Gamma^{'}}}$--${\rm J^{''}_{K_a^{''}\Gamma^{''}}}$)&(K)&(s$^{-1}$)& &(Debye$^{2}$)&(km s$^{-1}$)&(K km s$^{-1}$)  &($\tau$)\\
\hline
241.700&5(-0,5)--4(-0,4)E, $v_{t}$ = 0&47.93&6.04$\times$10$^{-5}$&11&16.15 &1.52$\pm$0.23&0.62$\pm$0.02&0.02 \\
241.767&5(1,5)--4(1,4)E, $v_{t}$ = 0  &40.39&5.81$\times$10$^{-5}$&11&15.53 &1.54$\pm$0.25 &0.65$\pm$0.01&0.03 \\
241.791&5(0,5)--4(0,4)A, $v_{t}$ = 0  &34.82&6.05$\times$10$^{-5}$&11&16.16 &1.50$\pm$0.32 &0.67$\pm$0.04&0.04 \\
241.806&5(4,2)--4(4,1)A, $v_{t}$ = 0  &115.16&2.18$\times$10$^{-5}$&11&5.82 &1.57$\pm$0.55 &0.22$\pm$0.06&0.02 \\
241.813&5(4,2)--4(4,1)E, $v_{t}$ = 0&122.72&2.18$\times$10$^{-5}$&11&5.82&1.72$\pm$0.62 &0.17$\pm$0.03&0.05\\
241.829&5(-4,1)--4(-4,0)E, $v_{t}$ = 0&130.82&2.19$\times$10$^{-5}$&11&5.85 &1.62$\pm$0.56 &0.12$\pm$0.02&0.03 \\
241.832&5(3,3)--4(3,2)A, $v_{t}$ = 0&84.62&3.87$\times$10$^{-5}$&11&10.33&1.52$\pm$0.26 &0.31$\pm$0.02&0.06\\
241.833&5(3,2)--4(3,1)A, $v_{t}$ = 0&84.62&3.87$\times$10$^{-5}$&11&10.33&1.52$\pm$0.26 &0.31$\pm$0.02&0.06\\
241.842&5(2,4)--4(2,3)A, $v_{t}$ = 0&72.53&5.12$\times$10$^{-5}$&11&13.66&1.56$\pm$0.52 &0.37$\pm$0.02&0.05\\
241.843&5(-3,3)--4(-3,2)E, $v_{t}$ = 0&82.53&3.88$\times$10$^{-5}$&11&10.36&1.56$\pm$0.52 &0.37$\pm$0.02&0.05\\
241.852&5(3,2)--4(3,1)E, $v_{t}$ = 0&97.53&3.89$\times$10$^{-5}$&11&10.40&1.52$\pm$0.24 &0.13$\pm$0.01&0.02 \\
241.879&5(-1,4)--4(-1,3)E, $v_{t}$ = 0&55.87&5.96$\times$10$^{-5}$&11&15.92&1.53$\pm$0.32 &0.42$\pm$0.06&0.03\\
241.887&5(2,3)--4(2,2)A, $v_{t}$ = 0&72.53&5.12$\times$10$^{-5}$&11&13.67&1.53$\pm$0.38 &0.35$\pm$0.02&0.02\\
241.904&5(2,3)--4(2,2)E, $v_{t}$ = 0&60.73&5.09$\times$10$^{-5}$&11&13.59&1.58$\pm$0.26 &0.78$\pm$0.02&0.03\\
241.904&5(-2,4)--4(-2,3)E, $v_{t}$ = 0&57.07&5.03$\times$10$^{-5}$&11&13.43&1.58$\pm$0.27 &0.78$\pm$0.02&0.03\\
\hline
\end{tabular}
\end{adjustbox}
\label{tab:MOLECULAR DATA}\\
\end{table*}

\begin{figure*}
\centering
\includegraphics[width=0.9\textwidth]{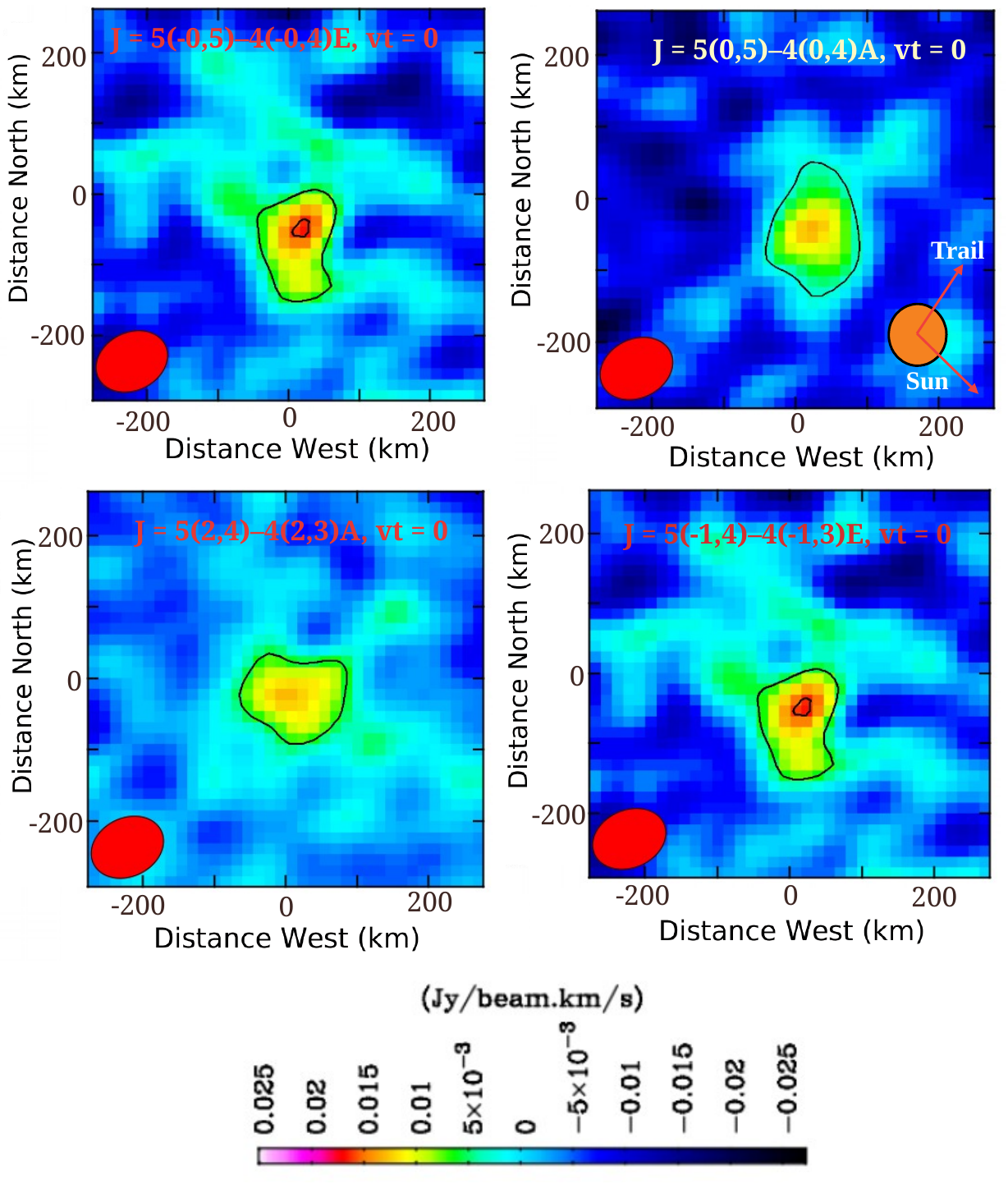}
\caption{Integrated intensity maps of \ce{CH3OH} towards comet ZTF. The red-coloured elliptical circles are the synthesized beam of the emission maps. The contour levels start at 3$\sigma$.}
\label{fig:emissionmap}
\end{figure*}

After detection of the emission lines of \ce{CH3OH}, we also made the integrated intensity maps of \ce{CH3OH} based on four high-intensity lines using the CASA task IMFIT. The integrated intensity maps of \ce{CH3OH} towards comet ZTF are shown in Figure~\ref{fig:emissionmap}. The integrated intensity maps clearly show that the \ce{CH3OH} emits from the inner coma region of comet ZTF. After fitting the 2D Gaussian over the integrated intensity maps, we observed that the emitting regions of \ce{CH3OH} vary between 0.47$^{\prime\prime}$ and 0.49$^{\prime\prime}$. The derived emitting regions of \ce{CH3OH} are lower than the synthesized beam sizes of the integrated intensity maps, which shows that the integrated intensity maps are not resolved.
	
\begin{table*}
\centering 
\caption{Production rate of \ce{CH3OH} towards comet ZTF.}
\begin{adjustbox}{width=0.7\textwidth}
\begin{tabular}{ccccccccccccccccccc}
\hline 
Frequency &Quantum number& $v_{exp}$&Production rate ($Q$)\\
(GHz)&(${\rm J^{'}_{K_a^{'}\Gamma^{'}}}$--${\rm J^{''}_{K_a^{''}\Gamma^{''}}}$)&(km s$^{-1}$)&(molecule s$^{-1}$)\\
\hline
241.700&5(-0,5)--4(-0,4)E, $v_{t}$ = 0&0.76$\pm$0.02&(7.25$\pm$0.29)$\times$10$^{26}$ \\
241.767&5(1,5)--4(1,4)E, $v_{t}$ = 0  &0.77$\pm$0.03&(6.82$\pm$0.62)$\times$10$^{26}$ \\
241.791&5(0,5)--4(0,4)A, $v_{t}$ = 0  &0.75$\pm$0.04&(7.02$\pm$0.55)$\times$10$^{26}$ \\
241.806&5(4,2)--4(4,1)A, $v_{t}$ = 0  &0.78$\pm$0.02&(7.36$\pm$0.78)$\times$10$^{26}$ \\
241.813&5(4,2)--4(4,1)E, $v_{t}$ = 0  &0.86$\pm$0.03&(7.55$\pm$0.21)$\times$10$^{26}$\\
241.829&5(-4,1)--4(-4,0)E, $v_{t}$ = 0&0.81$\pm$0.06&(7.82$\pm$0.30)$\times$10$^{26}$ \\
241.832&5(3,3)--4(3,2)A, $v_{t}$ = 0  &0.76$\pm$0.06&(6.98$\pm$0.53)$\times$10$^{26}$\\
241.833&5(3,2)--4(3,1)A, $v_{t}$ = 0&0.76$\pm$0.05&(6.97$\pm$0.55)$\times$10$^{26}$\\
241.842&5(2,4)--4(2,3)A, $v_{t}$ = 0&0.78$\pm$0.07&(7.09$\pm$1.20)$\times$10$^{26}$\\
241.843&5(-3,3)--4(-3,2)E, $v_{t}$ = 0&0.78$\pm$0.06&(7.10$\pm$1.22)$\times$10$^{26}$\\
241.852&5(3,2)--4(3,1)E, $v_{t}$ = 0&0.76$\pm$0.02&(7.52$\pm$0.98)$\times$10$^{26}$\\
241.879&5(-1,4)--4(-1,3)E, $v_{t}$ = 0&0.76$\pm$0.02&(7.36$\pm$0.76)$\times$10$^{26}$\\
241.887&5(2,3)--4(2,2)A, $v_{t}$ = 0&0.76$\pm$0.05&(7.29$\pm$0.90)$\times$10$^{26}$\\
241.904&5(2,3)--4(2,2)E, $v_{t}$ = 0&0.79$\pm$0.06&(7.82$\pm$0.55)$\times$10$^{26}$\\
241.904&5(-2,4)--4(-2,3)E, $v_{t}$ = 0&0.79$\pm$0.08&(7.92$\pm$0.22)$\times$10$^{26}$\\
\hline
&Average production rate ($Q_{CH_{3}OH}$)  & &(7.32$\pm$0.64)$\times$10$^{26}$   \\
\hline
\end{tabular}
\end{adjustbox}
\label{tab:production rate}\\
\end{table*}
	
\subsubsection{Production rate of \ce{CH3OH} towards comet ZTF}
To derive the production rate ($Q$) of \ce{CH3OH} using the line area ($\rm{\int T_{mb}dV}$) and rotational temperature towards comet ZTF, we used the following equation \citep{dra10}:
	
\begin{equation} \label{equation2}
Q = \frac{2}{\sqrt{\pi \ln 2}}\frac{k}{h}\frac{b\Delta v_{exp}}{DI(T)\nu}(\exp\left(\frac{h\nu}{kT}\right)-1)\rm{\int T_{mb}dV}
\end{equation}

In the above equation, $h$ and $K$ are the Planck and Boltzmann constants, $b$ is a dimensionless factor whose value is 1.22, $D$ is the diameter of the dish whose value is 12 m, $\Delta$ is the distance between Earth and comet, $I(T)$ indicates the integrated line intensity, which is defined in the CDMS line database, $v_{exp}$ is the half-width at half maximum (HWHM) of the emission lines of \ce{CH3OH}, $\nu$ is the rest frequency of the detected transition of \ce{CH3OH}, and $\rm{\int T_{mb}dV}$ is the integrated line area of molecule in km s$^{-1}$. For integrated line intensity values, we used $T$ = 75 K because that temperature value is very close to the derived rotational temperature (72$\pm$12 K) of \ce{CH3OH}. As per the CDMS molecular database, the value of $I(75 K)$ is 1.45$\times$10$^{-4}$ nm$^{2}$ MHz. The Equation~\ref{equation2} is well described in \citet{dra10}. Equation~\ref{equation2} is appropriate for estimating the production of \ce{CH3OH} towards comet ZTF because \ce{CH3OH} are found in LTE conditions and the radiated emission lines of \ce{CH3OH} are optically thin, which were emitted from the coma region. Using Equation~\ref{equation2}, we estimate the production rates of \ce{CH3OH} based on all detected transitions, which are listed in Table~\ref{tab:production rate}. The error bars in production rate values were estimated based on the error values of $v_{exp}$ and $\rm{\int T_{mb}dV}$. After averaging those 15 production rate values, we obtain the final production rate of \ce{CH3OH} towards comet ZTF is (7.32$\pm$0.64)$\times$10$^{26}$ molecules s$^{-1}$. The production rate of \ce{CH3OH} towards comet ZTF estimated by \cite{biv24} using the Haser model is 8.78$\times$10$^{26}$ molecules s$^{-1}$, which is very close to our estimated production rate of \ce{CH3OH} using Equation~\ref{equation2}. The abundance of \ce{CH3OH} with respect to \ce{H2O} towards comet ZTF is 1.52$\times$10$^{-2}$ (alternatively 1.52\%), where the production rate of \ce{H2O} towards comet ZTF is 4.8$\times$10$^{28}$ molecules s$^{-1}$ \citep{sc23}. 
	
\subsubsection{Comparision of the abundance of \ce{CH3OH} between ZTF and other comets}
Now, we compare the abundance of \ce{CH3OH} towards comet ZTF with other comets to understand the ice composition, as shown in Table~\ref{tab:abuncomp}. After comparison, we observe that the comet Austin (1990 V) exhibited a higher \ce{CH3OH} abundance. The abundance of \ce{CH3OH} towards comet ZTF is nearly similar to that of comets P/Halley, C/2002 T7 (LINEAR), and C/2012 K1 (PanSTARRS). The abundance of \ce{CH3OH} towards the ZTF is lower than that of P/Swift-Tuttle (1992), C/1996 B2 (Hyakutake), C/1995 O1 (Hale-Bopp), C/1999 H1 (Lee), 153P/2002 C1 (Ikeya-Zhang), C/2020 F3 (NEOWISE), 67P/Churyumov-Gerasimenko, and 46P/Wirtanen. Similarly, the abundance of \ce{CH3OH} towards ZTF is higher than that of Levy (1990 XX), C/2013 R1 (Lovejoy), C/2012 F6 (Lemmon), C/2012 S1 (ISON), and C/2021 A1 (Leonard). Since the abundance of \ce{CH3OH} towards ZTF is very similar to P/Halley, C/2002 T7 (LINEAR), and C/2012 K1 (PanSTARRS), there is a chance that the formation mechanism of \ce{CH3OH} and the icy compositions of these four comets may be the same.

\begin{table*}
\centering
\caption{Comparision of the abundance of \ce{CH3OH} with comet ZTF and other comets.}
\begin{tabular}{lcccc}
\hline
Comet & Abundance of \ce{CH3OH}& Reference \\
	& ($X = Q(CH_{3}OH)/Q(H_{2}O)$ in \%) & & \\
\hline
C/2023 E3 (ZTF)&1.52& This study\\
Austin (1990 V)&5&\cite{boc94a}\\
Levy (1990 XX)&0.91&\cite{boc94a}\\
P/Swift-Tuttle (1992)&4&\cite{boc94b}\\
P/Halley&1.71&\citet{eb94} \\
C/1996 B2 (Hyakutake)&2.01&\cite{bive99}\\
C/1995 O1(Hale-Bopp)&2.40&\cite{boc00}  \\
C/1999 H1 (Lee)&2.10&\cite{mu01}\\
153P/2002 C1 (Ikeya-Zhang)&2.50&\cite{di02} \\
C/2002 T7 (LINEAR)&1.50&\cite{rem08} \\
C/2013 R1 (Lovejoy)&0.37&\cite{ag14}  \\ 
C/2012 K1 (PanSTARRS)&1.33&\cite{co17}\\
C/2012 F6 (Lemmon)&0.96&\cite{bo17}  \\
C/2012 S1 (ISON)&0.48&\cite{bo17} \\
C/2020 F3 (NEOWISE)&2.30&\cite{biv22}\\
46P/Wirtanen&2.70&\cite{co23}\\
67P/Churyumov-Gerasimenko&2.10&\cite{biv23}\\
C/2021 A1 (Leonard)&0.88&\cite{biv24}\\
\hline
\end{tabular}
\label{tab:abuncomp}
\end{table*}
	
\subsubsection{Formation mechanism of \ce{CH3OH}) towards comet ZTF}
High-resolution ALMA observations show that \ce{CH3OH} emits from the very deep cometary coma, which means there is a chance to form this molecule in the grain surface. Previous studies showed that if \ce{CH3OH} is formed in cometary ices, then this molecule may be an old part of the star formation regions because comets move from several star-forming regions in the ISM \citep{rem08}. For the production of \ce{CH3OH}, two efficient reactions are available: (i) the radiative association of \ce{CH3}$^{+}$ and \ce{H2O} creates \ce{CH3OH2}$^{+}$ (\ce{CH3}$^{+}$ + \ce{H2O} $\rightarrow$ \ce{CH3OH2}$^{+}$) and \ce{CH3OH} is formed when \ce{CH3OH2}$^{+}$ recombine with an electron in the gas phase (\ce{CH3OH2}$^{+}$ + e$^{-}$ $\rightarrow$ \ce{CH3OH}), and (ii) the subsequential hydrogenation of CO formed \ce{CH3OH} in grain surface (CO + 2H $\rightarrow$ \ce{H2CO} + 2H $\rightarrow$ \ce{CH3OH}) \citep{rem08, gar19, fa23}. 
	
Now, we compare our estimated abundance of \ce{CH3OH} with the modelled value of \cite{gar19} to understand the possible formation pathway of \ce{CH3OH} towards comet ZTF. The estimated abundance of \ce{CH3OH} towards cometary atmosphere by \cite{gar19} is 1.3$\times$10$^{-2}$ (1.3\%) at 5$\times$10$^{9}$ yr (see Table~4 in \cite{gar19}), which was very close to the observed abundance of \ce{CH3OH} towards comet ZTF. The modelled value of \cite{gar19} is also very close to the abundance of \ce{CH3OH} towards other comets P/Halley, C/2002 T7 (LINEAR), and C/2012 K1 (PanSTARRS). That indicates \ce{CH3OH} may be formed via the subsequential hydrogenation of formaldehyde (\ce{H2CO}) on the grain surface of comets ZTF, P/Halley, C/2002 T7 (LINEAR), and C/2012 K1 (PanSTARRS). Similarly, the chemical models of \citet{gar19} show that \ce{CH3OH} is destroyed in the cometary atmosphere via photoionization processes (\ce{CH3OH} + h$\nu$ $\longrightarrow$ \ce{CH2} + \ce{H2O}).
	
\subsection{Results based on GMRT data}
After analysing the GMRT data, we did not detect the dust continuum emission from the comet ZTF. The upper limit flux density of comet ZTF at frequency 1250 MHz was $\leq$1.50$\pm$0.21 mJy, where the synthesized beam size of the continuum emission image was 2.66$^{\prime\prime}\times$1.50$^{\prime\prime}$. After that, we extracted the atomic spectra by making a 5.0$^{\prime\prime}$ diameter circular region over the spectra images. After analysing the atomic spectra, we could not detect the HI line at 1420 MHz. We estimate the upper limit column density of atomic HI using the following equation \citep{chen13}:
	
\begin{equation} \label{equation4}
N_{HI} = 1.823\times10^{18}\rm{\int T_{s} \tau dV}
\end{equation}
In the above equation, $T_{s}$ indicates the spin temperature of HI in K, $\tau$ is the optical depth of the HI spectra, and $\int$dv is the integrated area in km s$^{-1}$. The equation~\ref{equation4} is well described in \cite{pal24}. We used the value of $T_{s}$ = 100 K \citep{pa17}, $\tau$ $\leq$ 0.23, and $\int$dv $\leq$ 0.36. Using Equation~\ref{equation4}, we found the upper limit column density of atomic HI is $\leq$1.50$\times$10$^{19}$ cm$^{-2}$.
	
\section{Conclusion} 
\label{conclu}
In this article, we present the radio and millimeter-wavelength studies of comet ZTF using the GMRT band L and the ALMA band 6. The principal conclusions derived from this study are summarized below:\\\\
1. We detected the dust continuum emission from comet ZTF using the ALMA between the frequency ranges of 223.85 GHz and 243.55 GHz. The activity of dust production (Af$\rho$) of comet ZTF is 2280$\pm$50 cm. \\\\
2. We detected the rotational emission lines of \ce{CH3OH} towards comet ZTF using the ALMA. A total of 15 transition lines of \ce{CH3OH} is detected. Using the LTE model, we found the column density and excitation temperature of \ce{CH3OH} are (4.50$\pm$0.25)$\times$10$^{14}$ cm$^{-2}$ and 70$\pm$3 K. From the integrated emission maps, we observed that the emission lines of \ce{CH3OH} emit from the coma region of the comet ZTF. The production rate of \ce{CH3OH} on March 2, 2023 is (7.32$\pm$0.64)$\times$10$^{26}$ molecules s$^{-1}$. The abundance of \ce{CH3OH} with respect to \ce{H2O} in the coma of the comet is 1.52\%. We compared our derived abundance of \ce{CH3OH} with the modelled value of \cite{gar19}, and we noticed that the observed and modelled values are nearly similar. We claim that \ce{CH3OH} is formed via the subsequential hydrogenation of formaldehyde (\ce{H2CO}) on the grain surface of comet ZTF.\\\\
3. From the GMRT data, continuum emission and emission line of atomic HI are not detected. The upper limit flux density of comet ZTF at frequency 1250 MHz is $\leq$1.50$\pm$0.21 mJy. The upper limit column density of atomic HI towards comet ZTF is $\leq$1.50$\times$10$^{19}$ cm$^{-2}$.
	
\section*{acknowledgments}
We thank the anonymous reviewers for their helpful comments, which helped to make the manuscript stronger. A.M. acknowledges the Swami Vivekananda Merit cum Means Scholarship (SVMCM), Government of West Bengal, for financial support for this research. Supplementary materials, including additional plots and data, are available from the corresponding author upon reasonable request. We are grateful to the staff of the Giant Metrewave Radio Telescope (GMRT) for their support during the observation. The raw data of comet ZTF presented in this paper can be accessed through the GMRT archive using the proposal code ddtC267. The GMRT is operated by the National Centre for Radio Astrophysics (NCRA), a facility of the Tata Institute of Fundamental Research (TIFR). This paper makes use of the following ALMA data: ADS /JAO.ALMA\#2022.1.00997.T. ALMA is a partnership of ESO (representing its member states), NSF (USA), and NINS (Japan), together with NRC (Canada), MOST and ASIAA (Taiwan), and KASI (Republic of Korea), in co-operation with the Republic of Chile. The joint ALMA Observatory is operated by ESO, AUI/NRAO, and NAOJ. We also acknowledge Gideon van Buitenen (\url{http://astro.vanbuitenen.nl/comet/2022E3}) for providing the track of comets and images of the inner solar system.

\section*{Conflicts of interest}
The authors declare no conflict of interest.

\bibliographystyle{aasjournal}

\end{document}